\documentclass{amsart}%
\usepackage{amsfonts}
\usepackage{amsmath}
\usepackage{amssymb}
\usepackage{graphicx}%
\providecommand{\U}[1]{\protect\rule{.1in}{.1in}}
\theoremstyle{plain}
\newtheorem{theorem}{Theorem}[section]

\newtheorem{proposition}[theorem]{Proposition}
\theoremstyle{definition}

\theoremstyle{remark}

\numberwithin{equation}{section}

\begin{document}
\title[The Calogero type integrable discretizations]{On the completely integrable Calogero type discretizations of nonlinear Lax
integrable dynamical systems and the related Markov type co-adjoint orbits}
\author{Anatolij K. Prykarpatski}
\address{The Department of Applied Mathematics at AGH University of Science and
Technology, Krakow 30059, Poland,}
\email{pryk.anat@ua.fm, prykanat@cybergal.com}
\date{present}

\begin{abstract}
The Calogero type matrix discretization scheme is applied to constructing the
Lax type integrable discretizations of one wide enough class of nonlinear
integrable dynamical systems on functional manifolds. Their Lie-algebraic
structure and complete integrability related with co-adjoint orbits on the
Markov co-algebras is discussed. It is shown that a set of conservation laws
and the associated Poisson structure ensue as a byproduct of the approach
devised. Based on the \ Lie algebras quasi-representation property the
limiting procedure of finding the nonlinear dynamical systems on the
corresponding functional spaces is demonstrated.

\end{abstract}
\maketitle

\section{\bigskip Introduction: the discretization and related Markov algebra
splitting}

With a fairly generous definition, a one-dimensional real-valued discrete
nonlinear dynamical system on a manifold $M\subset l_{2}(\mathbb{Z}%
;\mathbb{R}^{m})$ for some\ \ finite $\ \ m\in\mathbb{Z}_{+}$ \ \ is any
evolution equation that can be written down as%

\begin{equation}
du/dt=K[u], \label{S1}%
\end{equation}
where $t\in\mathbb{R}$ is the evolution parameter, $\ \ u\in M\ $ and
$K:M\rightarrow T(M)$ \ is some smooth enough vector field \cite{AbMa,Arno} on
the \ manifold $M.$ \ Very often such equations \ (\ref{S1}) can be naturally
obtained as the standard discretization \cite{AbLa,CaDe,Kupe,Newe} of a given
smooth nonlinear differential dynamical system
\begin{equation}
d\mathrm{u}/dt=\mathcal{K}[\mathrm{u}] \label{S2}%
\end{equation}
on a functional submanifold $\mathcal{M}\subset L_{2}(\mathbb{R}%
;\mathbb{R}^{m}),$ generated by a smooth vector field $\mathcal{K}%
:\mathcal{M}\rightarrow T(\mathcal{M}).$ Namely, there exist such mesh points
$x_{j}\neq x_{i}\in\mathbb{R}\ $\ for $i\neq j\in\mathbb{Z},$ that the
corresponding vector $\{\mathrm{u}(x_{j})\in\mathbb{R}^{m}:j\in\mathbb{Z}\}$
$=u\in M\ $\ and the suitable discretization of \ (\ref{S2}) coincides with
\ (\ref{S1}).

Other approach to the discretization of \ (\ref{S2}) is based on the Calogero
type \cite{CaFr,CaDe} scheme of constructing finite-dimensional
quasi-representations of the infinite-dimensional Heisenberg-Weyl algebra of
operators $\mathfrak{h:=}\{\hat{x},D_{x},\hat{1}:x\in\mathbb{R}\},$ where
$D_{x}:=\partial/\partial x,$ in some functional submanifold $\mathcal{M}_{0}$
$\subset C^{\infty}(\mathbb{[}a,b];\mathbb{R})$ $\cap L_{2}(\mathbb{[}%
a,b];\mathbb{R})$ of differentiable functions, as owing to the well known
von-Neumann theorem \cite{Neum,Zeid}, there exists no exact representation of
$\mathfrak{h}$ in a finite-dimensional functional subspace $\mathcal{M}%
_{0}^{N}\subset\mathcal{M}_{0}$ for any $N\in\mathbb{Z}_{+}.$ \ For example,
any smooth scalar function $f\in\mathcal{M}_{0}$ on an interval $[a,b]\subset
\mathbb{R}$ can be interpolated \cite{Zeid,CaFr} in a polynomial form as
follows:%
\begin{align}
f(x)  &  \rightarrow f_{N}(x):=\sum_{j=1}^{N}(f(x_{j})\rho_{j}^{-1}%
)e_{j}(x),\text{ \ \ \ \ \ }\label{S3}\\
e_{j}(x)  &  :=%
{\displaystyle\prod\limits_{i=\overline{1,N},i\neq j}}
(x-x_{i}),\rho_{j}:=%
{\displaystyle\prod\limits_{i=\overline{1,N},i\neq j}^{N}}
(x_{j}-x_{i})\nonumber
\end{align}
and its derivative, respectively, as
\begin{equation}
D_{x}f(x)\rightarrow D_{x}f_{N}(x)=\sum_{i,j=\overline{1,N}}Z_{ij}%
(f(x_{j})\rho_{j}^{-1})\ e_{i}(x)),\text{ \ \ } \label{S4}%
\end{equation}
where $f_{N}(x)\in\mathcal{M}_{0}^{N}$ subject to the polynomial basis
$\{e_{j}(x)\in\mathcal{M}_{0}^{N}:j=\overline{1,N}\}.$ \ Then the well known
Calogero type quasi-representation \ \cite{CaFr,LuJaKuPr,Lebe,Cont,CoGrMo} of
the Heisenberg-Weyl algebra $\mathfrak{h}$ is obtained as
\begin{align}
End(\mathcal{M}_{0})  &  \ni\hat{x}\rightarrow X:=diag\{x_{1},x_{2}%
,...,x_{N}\}\in End\text{ }l_{2}(\mathbb{Z}_{N};\mathbb{R}),\label{S5}\\
End(\mathcal{M}_{0})  &  \ni\hat{1}\rightarrow I:=diag\{1,1,...,1\}\in
End\text{ }l_{2}(\mathbb{Z}_{N};\mathbb{R})\nonumber\\
End(\mathcal{M}_{0})  &  \ni D_{x}\rightarrow Z:=\{Z_{ij}:=(x_{i}-x_{j}%
)^{-1},i\neq j=\overline{1,N};\nonumber\\
Z_{ii}  &  :=\sum_{j=1,j\neq i}^{N}(x_{i}-x_{j})^{-1}:i=\overline{1,N}\}\in
End\text{ }l_{2}(\mathbb{Z}_{N};\mathbb{R}),\nonumber
\end{align}
where \ interpolating mesh points $x_{i}\neq x_{j}\in\mathbb{R},i\neq
j=\overline{1,N},$ are chosen to be different and satisfying in a suitably
defined finite-dimensional Hilbert space $l_{2}(\mathbb{Z}_{N};\mathbb{R})$
the strong as $N\rightarrow\infty$ \ limiting canonical Lie algebra
relationship%
\begin{equation}
\lim_{N\rightarrow\infty}([Z,X]-I\mathcal{)}=0. \label{S6}%
\end{equation}

The matrix quasi-representations \ (\ref{S5}) make it possible to construct
easily a \textit{naive} matrix discretization of the nonlinear dynamical
system \ (\ref{S2}) as\ follows:
\begin{equation}
dU^{(m)}/dt=\mathcal{K}(U^{(m)},[Z^{(m)},U^{(m)}],[Z^{(m)},[Z^{(m)}%
,U^{(m)}]],...\underset{(p\text{-}times)}{[Z^{(m)},...,[Z^{(m)},U^{(m)}]]}),
\label{S7}%
\end{equation}
where the matrices
\begin{align}
U^{(m)}  &  :=u(X^{(m)})=diag(u(x_{1}),u(x_{2}),...,u(x_{N}))\in End\text{
}l_{2}(\mathbb{Z}_{N};\mathbb{R})^{\otimes m},\label{S8}\\
& \nonumber\\
Z^{(m)}  &  :=\underset{(m\text{-}times)}{Z\otimes Z\otimes...\otimes Z\text{
}}\in End\text{ }l_{2}(\mathbb{Z}_{N};\mathbb{R})^{\otimes m}\nonumber
\end{align}
belong to the tensor product matrix space%

\begin{equation}
Endl_{2}(\mathbb{Z}_{N};\mathbb{R})^{\otimes m}:=Endl_{2}(\mathbb{Z}%
_{N};\mathbb{R})\otimes Endl_{2}(\mathbb{Z}_{N};\mathbb{R})\underset
{(m\text{-}times)}{\otimes...\otimes}Endl_{2}(\mathbb{Z}_{N};\mathbb{R}%
).\label{S9}%
\end{equation}
When deriving \ the matrix equation (\ref{S7}), \ we \ took \ \ into
\ \ account \ that $\ \ \mathcal{K}[\mathrm{{u}}]\mathrm{:=\mathcal{K}%
\emph{(}u\emph{,}}D_{x}\mathrm{{u},}D_{x}^{2}\mathrm{{u},...}D_{x}%
^{p}\mathrm{{u}})$\ \ for some fixed $p\in\mathbb{Z}_{+},$ and for arbitrary
operator mapping $\varphi_{N}(\hat{x}):\mathcal{M}_{0}^{N}\rightarrow
\mathcal{M}_{0}^{N}$ we made use of the Calogero type quasi-representation
property:
\begin{equation}
End(\mathcal{M}_{0})\ni(D_{x}^{n}\varphi_{N})(\hat{x})\rightarrow
\underset{(n\text{-}times)}{[Z,[Z,[Z,}...,[Z,\varphi(X)]]]...]\in End\text{
}l_{2}(\mathbb{Z}_{N};\mathbb{R}),\label{S10}%
\end{equation}
which holds for arbitrary operator derivatives $\ (D_{x}^{n}\varphi_{N}%
)(\hat{x}):\mathcal{M}_{0}^{N}\rightarrow\mathcal{M}_{0}^{N},n\in
\mathbb{Z}_{+}.$

\ Now we will take into account the observation \cite{Meno} that the
quasi-representations \ (\ref{S5}) belong, respectively, \ to the Markov
direct sum splitting of the general Lie algebra $\mathfrak{\ }gl(N;\mathbb{R}%
):=\mathfrak{g}=\mathrm{M}(\mathfrak{g})\oplus\mathrm{E}(\mathfrak{g}):$%
\begin{equation}
I,X\in\mathrm{E}(\mathfrak{g}),\text{ \ \ }Z\in\mathrm{M}(\mathfrak{g}),
\label{M1.21}%
\end{equation}
where, by definition, the linear subspaces
\begin{align}
\mathrm{M}(\mathfrak{g})  &  :=\{\mathrm{M}(A)\in\mathfrak{g}:\mathrm{M}%
(A)=A-diag(eA)\}\label{M1.22}\\
\mathrm{E}(\mathfrak{g})  &  :=\{\mathrm{E}(A)\in\mathfrak{g}:\mathrm{E}%
(A)=diag(eA),\text{ }e:=(1,1,...,1)\in l_{2}(\mathbb{Z}_{N};\mathbb{R})^{\ast
},\nonumber
\end{align}
are Lie subalgebras of the Lie algebra $\mathfrak{g}.$ Introduce now, by
definition, projections $\mathrm{P}_{\mathrm{M}(\mathfrak{g})}:\mathfrak{g}%
=\mathrm{M}(\mathfrak{g})\subset\mathfrak{g,}$ $\mathrm{P}_{\mathrm{E}%
(\mathfrak{g})}:\mathfrak{g}=\mathrm{E}(\mathfrak{g})\subset\mathfrak{g.}$
\ Then within the standard $\mathrm{R}$-matrix approach
\cite{Blas,BlPrSa,FaTa,PrMy,ReS-T} the expression
\begin{equation}
\lbrack X,Y]_{\mathrm{R}}:=\ [\mathrm{R}X,Y]+[X,\mathrm{R}Y],\text{
}\mathrm{R}:=1/2(\mathrm{P}_{\mathrm{M}(\mathfrak{g})}-\mathrm{P}%
_{\mathrm{E}(\mathfrak{g})}), \label{M1.23}%
\end{equation}
for arbitrary $X,Y\in\mathfrak{g}$ \ defines on $\mathfrak{g}$ a new Lie
commutator structure, generating on the space $\mathcal{D}(\mathfrak{g})$ the
deformed Lie-Poisson bracket%
\begin{equation}
\{\gamma,\eta\}_{\mathrm{R}}(\alpha):=<\alpha,[\nabla\gamma(\alpha),\nabla
\eta(\alpha)]_{\mathrm{R}}>=<\alpha,[\mathrm{P}_{\mathrm{M}(\mathfrak{g}%
)}\nabla\gamma(\alpha),\mathrm{P}_{\mathrm{M}(\mathfrak{g})}\nabla\eta
(\alpha)]> \label{M1.24}%
\end{equation}
for $\gamma,\eta\in\mathcal{D}(\mathfrak{g})$ and any $\alpha\in
\mathfrak{g}^{\ast},$ generalizing the classical Lie-Poisson bracket
\begin{equation}
\{\gamma,\eta\}(\alpha):=<\alpha,[\nabla\gamma(\alpha),\nabla\eta
(\alpha)]>=<\alpha,[\nabla\gamma(\alpha),\nabla\eta(\alpha)]> \label{M1.24a}%
\end{equation}
on $\mathfrak{g.}$ Here the bi-linear trace-functional on the $\mathfrak{g}$
\
\begin{equation}
<X,Y>:=\mathrm{tr}\ (XY)\ \label{M1.25}%
\end{equation}
for $X,Y\in\mathfrak{g}$ is nondegenerate and $Ad$-invariant. Taking into
account that with respect to this trace-functional \ (\ref{M1.25}) the Lie
algebra $\mathfrak{g\simeq g}^{\ast},$ the Poisson bracket \ (\ref{M1.24})
generates for any Hamiltonian function $H\in\mathcal{D}(\mathfrak{g})$ the
following dynamical system on arbitrary $\alpha\in\mathfrak{g:}$%
\begin{equation}
d\alpha/dt=\mathrm{P}_{\mathrm{E}(\mathfrak{g})^{\bot}}[\mathrm{P}%
_{\mathrm{M}(\mathfrak{g})}\nabla H(\alpha),\alpha], \label{M1.26}%
\end{equation}
where we took into account that projections $\mathrm{P}_{\mathrm{M}%
(\mathfrak{g})}^{\ast}\simeq\mathrm{P}_{\mathrm{E}(\mathfrak{g})^{\bot}}$ and
$\mathrm{P}_{\mathrm{E}(\mathfrak{g})}^{\ast}\simeq\mathrm{P}_{\mathrm{M}%
(\mathfrak{g}\mathcal{)}^{\bot}}.$ This construction becomes more simpler in
the case when the Hamiltonian function $H\in\mathrm{I}(\mathfrak{g})$ is taken
to be a Casimir one with respect to the classical Lie-Poisson bracket
\ (\ref{M1.24a}), satisfying the condition
\begin{equation}
\lbrack\alpha,\nabla H(\alpha)]=0 \label{M1.27}%
\end{equation}
for any $\alpha\in\mathfrak{g.}$

Amongst nonlinear differential dynamical systems \ (\ref{S2}) there exist a
wide class of nonlinear evolution equations which are Lax type
\cite{CaDe,Newe,Novi,FaTa} integrable and whose discretizations are often very
important for their numerical analysis and diverse applications. \ Yet in
general, the presented above directly discretized matrix dynamical system
\ (\ref{S7}) does not \textit{a priori} inherits the Lax type
integrability\ of \ (\ref{S2}). Thus, a natural question arises: \textit{how
to construct a priori Lax type integrable matrix discretization of a given Lax
type integrable nonlinear dynamical system \ (\ref{S2})? }

As the Lax type representations of the presumably integrable dynamical systems
\ (\ref{S2}) depend on an arbitrary spectral parameter $\lambda\in\mathbb{C},$
it motivates us to study their corresponding matrix Lax type
quasi-representations, also depending on the spectral parameter $\lambda
\in\mathbb{C},$ as well as depending on the basis matrix representation
operators \ (\ref{S5}), belonging, respectively, \ to the Markov direct sum
splitting of the general Lie algebra $\mathfrak{\ }gl(N;\mathbb{R}%
):=\mathfrak{g}=\mathrm{M}(\mathfrak{g})\oplus\mathrm{E}(\mathfrak{g}).$ This
can be done effectively by means of introducing the notion of the metrized
loop \cite{Blas,FaTa,PrMy,BlPrSa} algebra $\mathfrak{\tilde{g}:=g\otimes
}\mathbb{C[}\mathfrak{[\lambda},$ $\lambda^{-1}]]$ and the related Lax type
integrable Poisson flows on \ it.

Below I present a solution to this posed above question in the case of a
special class of nonlinear Lax type integrable dynamical systems on functional
manifolds making use of the Calogero type discretization scheme and the
analysis of the Markov type co-adjoint orbits by means of the related
Lie-algebraic techniques.

\section{The Lie algebraic setting and the Calogero type linear matrix
spectral problems}

There is introduced a metrized loop algebra $\mathfrak{\tilde{g}:=g\otimes
}\mathbb{C[}\mathfrak{[\lambda},$ $\lambda^{-1}]],$ generated by the Lie
algebra$\ \mathfrak{g:}=gl(N;\mathbb{R})$ and the related Laurent series%
\begin{equation}
\mathfrak{\tilde{g}}:\mathcal{=}\{X(\lambda)=\sum_{j\ll\infty}X_{j}\lambda
^{j}:X_{j}\in\mathfrak{g},\text{ \ }\mathbb{Z}\ni j\ll\infty,\lambda
\in\mathbb{C}\}. \label{M1.1}%
\end{equation}
It is endowed with the standard matrix Lie bracket
\begin{equation}
\lbrack X(\lambda),Y(\lambda)]:=\sum_{s\ll\infty}\lambda^{s}(\sum
_{j+k=s}[X_{j},Y_{k}]), \label{M1.2}%
\end{equation}
defined for any $X(\lambda),Y(\lambda)\in\mathfrak{\tilde{g}}$ \ and the
simplest $Ad$-invariant nondegenerate scalar product%
\begin{equation}
<X(\lambda),Y(\lambda)>_{-1}=\mathrm{Tr}(X(\lambda)Y(\lambda))\ :=res\text{
}tr(X(\lambda)Y(\lambda)), \label{M1.3}%
\end{equation}
satisfying for all $X(\lambda),Y(\lambda)$ and $Z(\lambda)$ $\in
\mathfrak{\tilde{g}}$ \ the condition%
\begin{equation}
<X(\lambda),[Y(\lambda),Z(\lambda)]>_{-1}=<[X(\lambda),Y(\lambda
)],Z(\lambda)>_{-1}. \label{M1.4}%
\end{equation}
Consider now the introduced above \ Markov splitting (\ref{M1.22}) of the Lie
algebra $\mathfrak{g:}$ $\ $%
\begin{equation}
\mathfrak{g}:=\mathrm{M}(\mathfrak{g})\oplus\mathrm{E}(\mathfrak{g}%
\mathcal{)}, \label{M1.4a}%
\end{equation}
where for any $X\in\mathfrak{g}$
\begin{equation}
\mathrm{M}(X):=X-diag(eX),\text{ \ \ }\mathrm{E}(X):=diag(eX),\text{
}e:=(1,1,...,1)\in l_{2}(\mathbb{Z}_{N})^{\ast}, \label{M1.4b}%
\end{equation}
and the components $\mathrm{M}(\mathfrak{g}\mathcal{)}\mathcal{\subset}$
$\mathfrak{g}$ and $\mathrm{E}(\mathfrak{g)}\subset$ $\mathfrak{g}$ \ are
suitable matrix Lie subalgebras, that is $[\mathrm{M}(\mathfrak{g}%
\mathcal{)},\mathrm{M}(\mathfrak{g}\mathcal{)}]\subset\mathrm{M}%
(\mathfrak{g}\mathcal{)}$ and $[\mathrm{E}(\mathfrak{g)},\mathrm{E}%
(\mathfrak{g)}]=0\in\mathrm{E}(\mathfrak{g)}.$

The following observation is crucial for our next analysis: the loop algebra
$\mathfrak{\tilde{g}}$ \ inherits the Markov type splitting \ (\ref{M1.4a})
into two Lie subalgebras:
\begin{equation}
\mathfrak{\tilde{g}}=\mathfrak{\tilde{g}}_{+}\oplus\mathfrak{\tilde{g}}_{-},
\label{M1.4c}%
\end{equation}
where $\ $the projection $\mathcal{\ }P_{+}\mathfrak{\tilde{g}}:\mathcal{=}%
\mathfrak{\tilde{g}}_{+},$
\begin{equation}
\mathfrak{\tilde{g}}_{+}:=\{X(\lambda)=\sum_{j\in\mathbb{Z}_{+}}X_{j}%
\lambda^{j}:X_{0}\in\mathrm{M}(\mathfrak{g}\mathcal{)},\text{ \ }X_{j}%
\in\mathfrak{g},\text{ \ }\mathbb{N}\ni j\ll\infty,\lambda\in\mathbb{C}%
\}\ \label{M1.5}%
\end{equation}
and \ the projection $\mathcal{\ }\mathrm{P}_{-}\mathfrak{\tilde{g}%
}:\mathcal{=}\mathfrak{\tilde{g}}_{-},$
\begin{equation}
\mathfrak{\tilde{g}}_{-}:=\{X(\lambda)=\sum_{j\in\mathbb{Z}_{-}}Y_{j}%
\lambda^{j}:Y_{0}\in\mathrm{E}(\mathfrak{g)},\text{ \ }Y_{j}\in\mathfrak{g}%
,\text{ \ }j\in\mathbb{Z}_{-}\backslash\{0\}\ ,\lambda\in\mathbb{C}%
\}\ \label{M1.6}%
\end{equation}
satisfy the Lie commutator relationships $[\mathfrak{\tilde{g}}_{+}%
,\mathfrak{\tilde{g}}_{+}]\subset\mathfrak{\tilde{g}}_{+}$ and
$[\mathfrak{\tilde{g}}_{-},\mathfrak{\tilde{g}}_{-}]\subset\mathfrak{\tilde
{g}}_{-}.$ It is easy to calculate the adjoint spaces $\ \mathfrak{\tilde{g}%
}_{+}^{\ast}$ and $\ \mathfrak{\tilde{g}}_{-}^{\ast}$ \ to Lie subalgebras
\ (\ref{M1.5}) and \ (\ref{M1.6}):
\begin{align}
\mathfrak{\tilde{g}}_{+}^{\ast}  &  \simeq\mathfrak{\tilde{g}}_{-}^{\intercal
}=\{Y(\lambda)=\sum_{j\in\mathbb{Z}_{-}}Y_{j}\lambda^{j}:Y_{0}\in
\mathrm{E}(\mathfrak{g)}^{\bot},\text{ \ }Y_{j}\in\mathfrak{g},\text{
\ }\ j\in\mathbb{Z}_{-}\backslash\{0\}\ ,\lambda\in\mathbb{C}\},\label{M1.7}\\
& \nonumber\\
\mathfrak{\tilde{g}}_{-}^{\ast}  &  \simeq\mathfrak{\tilde{g}}_{+}^{\intercal
}=\{X(\lambda)=\sum_{j\in\mathbb{Z}_{+}}X_{j}\lambda^{j}:X_{0}\in
\mathrm{M}(\mathfrak{g}\mathcal{)}^{\bot},\text{ \ }X_{j}\in\mathfrak{g}%
,\text{ \ }\mathbb{N}\ni j\ll\infty,\lambda\in\mathbb{C}\},\nonumber
\end{align}
where we took into account that
\begin{align}
\mathrm{M}(\mathfrak{g}\mathcal{)}^{\ast}  &  \simeq\mathrm{E}(\mathfrak{g}%
\mathcal{)}^{\bot}=\{Y\in\mathfrak{g}:\text{ }diag(Y)=0\},\label{M1.7a}\\
& \nonumber\\
\mathrm{E}(\mathfrak{g}\mathcal{)}^{\ast}  &  \simeq\mathrm{M}(\mathfrak{g}%
\mathcal{)}^{\bot}=\{X\in\mathfrak{g}:\text{ }\ X\ =q\otimes e,\nonumber\\
& \nonumber\\
e  &  :=(1,1,...,1)\ \in l_{2}(\mathbb{Z}_{N};\mathbb{R})^{\ast},q\in
l_{2}(\mathbb{Z}_{N};\mathbb{R})\}.\nonumber
\end{align}
The splitting \ (\ref{M1.4c}) makes it possible to define a related classical
$\mathrm{R}$-structure on the loop algebra $\mathfrak{\tilde{g}}:$ for any
$\ X(\lambda),Y(\lambda)\in\mathfrak{\tilde{g}}$ \ the commutator
\begin{equation}
\lbrack X(\lambda),Y(\lambda)]_{\mathrm{R}}:=\ \ ([\mathrm{R}X(\lambda
),Y(\lambda)]+\ [X(\lambda),\mathrm{R}Y(\lambda)]), \label{M1.8}%
\end{equation}
satisfies the Lie algebra\ commutator property, where the linear space
homomorphism $\mathrm{R}:\mathfrak{\tilde{g}\rightarrow}$\ $\mathfrak{\tilde
{g}}$ \ is defined for an arbitrary $X(\lambda)\in\mathfrak{\tilde{g}}$ as
\begin{equation}
\mathrm{R}X(\lambda):=\ 1/2(\mathrm{P}_{+}X(\lambda)-\mathrm{P}_{-}%
X(\lambda)). \label{M1.9}%
\end{equation}
The following important classical (see, for instance, \cite{Blas,BlPrSa,PrMy})
theorem holds.

\begin{theorem}
\label{Tm_M2.1} (Adler-Kostant-Souriau) \ Let smooth functionals $\gamma
,\eta:\mathfrak{\tilde{g}}^{\ast}\rightarrow\mathbb{R}$ \ \ be Casimir ones
subject to the Lie bracket \ (\ref{M1.2}), \ that is
\begin{equation}
\lbrack\nabla\gamma(l(\lambda)),l(\lambda)]=0=[\nabla\eta(l(\lambda
)),l(\lambda)]. \label{M1.10a}%
\end{equation}
for any $l(\lambda)\in\mathfrak{g}^{\ast}.$ Then their modified Lie-Poisson
bracket
\begin{equation}
\{\gamma,\eta\}\ :=<l(\lambda),[\nabla\gamma(l(\lambda)),\nabla\eta
(l(\lambda))]_{\mathrm{R}}>_{-1} \label{M1.10b}%
\end{equation}
\bigskip vanishes: $\ \{\gamma,\eta\}\ =0$ on the whole space
$\mathfrak{\tilde{g}}^{\ast}.$
\end{theorem}

Based on the splitting \ (\ref{M1.4c}) one can easily to calculate the actions
of adjoint operators $\ \mathrm{P}_{+}^{\ast}:\ \mathfrak{\tilde{g}%
}\rightarrow\mathfrak{\tilde{g}}\mathcal{_{-}^{\intercal}}\subset
\mathfrak{\tilde{g}}_{-}$ \ and $\mathrm{P}_{-}^{\ast}:\mathfrak{\tilde{g}%
}\rightarrow\mathfrak{\tilde{g}}\mathcal{_{+}^{\intercal}}\subset
\mathfrak{\tilde{g}}_{+}.$ Namely, \ owing to the identification
$\mathfrak{\tilde{g}}\simeq\mathfrak{\tilde{g}}^{\ast}$ one $\ $finds that the
following equalities $\ $%
\begin{equation}
\mathrm{P}_{+}^{\ast}=\mathrm{P}_{\mathfrak{\tilde{g}}_{-}^{\intercal}%
}:\mathfrak{\tilde{g}}\rightarrow\mathfrak{\tilde{g}}\mathcal{_{-}^{\intercal
}}\subset\mathfrak{\tilde{g}}_{-}\ \label{M1.11a}%
\end{equation}
and
\begin{equation}
\mathrm{P}_{-}^{\ast}=\mathrm{P}_{\mathfrak{\tilde{g}}_{+}^{\intercal}%
}:\mathfrak{\tilde{g}}\rightarrow\mathfrak{\tilde{g}}\mathcal{_{+}^{\intercal
}}\subset\mathfrak{\tilde{g}}_{+} \label{M1.11b}%
\end{equation}
\ hold.

Theorem \ \ref{Tm_M2.1} above and the equalities (\ref{M1.11a}) and
\ (\ref{M1.11b}) make it possible to construct a wide class of Liouvile
integrable dynamical systems \cite{BlPrSa,PrMy} on the Markov matrix subspace
$\mathrm{E}(\mathfrak{g}),$ if to reduce the related Hamiltonian vector field
\begin{equation}
\frac{d}{dt}\alpha(\lambda):=\{H,\alpha(\lambda\}=[\mathrm{P}_{+}\nabla
H(\alpha(\lambda)),\alpha(\lambda)],\text{ \ } \label{M1.12}%
\end{equation}
on the element $\alpha(\lambda)\in\mathfrak{\tilde{g}\simeq\tilde{g}}^{\ast},$
generated by a specially chosen smooth Casimir functional
$H\ \ :\mathfrak{\tilde{g}}\rightarrow\mathbb{R}.$ The latter is considered
with respect to the standard Lie-Poisson bracket%
\begin{equation}
\{\gamma,\eta\}_{Lie}\ :=<l(\lambda),[\nabla\gamma(l(\lambda)),\nabla
\eta(l(\lambda))]\ >_{-1} \label{M1.13}%
\end{equation}
for any smooth functionals $\gamma,\eta\in\mathcal{D}(\mathfrak{\tilde{g}}).$

Consider the following smooth functionals $\gamma_{n}^{(k)}:\mathfrak{\tilde
{g}}^{\ast}\rightarrow\mathbb{R},$ $n,k\in\mathbb{Z}_{+},$ where
\begin{align}
\gamma_{n}^{(k)}\  &  :=1/(n+1)<(\alpha(\lambda)\lambda^{-|\alpha(\lambda
)|})^{n+1},\lambda^{k+|\alpha(\lambda)|}>_{-1}=\label{M2.1}\\
&  =1/(n+1)Tr(\lambda^{k+|\alpha(\lambda)|}\ (\alpha(\lambda)\lambda
^{-|\alpha(\lambda)|})^{n}).\nonumber
\end{align}
They are, evidently, Casimir ones for the Poisson bracket (\ref{M1.13}), whose
gradients equal
\begin{equation}
\nabla\gamma_{n}^{(k)}(\alpha(\lambda))=(\alpha(\lambda)\lambda^{-|\alpha
(\lambda)|})^{n}\lambda^{k}.\ \label{M2.2}%
\end{equation}
Taking, for example, $n=2$ and $k=4,$ the corresponding value of the element
$\mathrm{P}_{+}\nabla\gamma_{2}^{(4)}(\alpha(\lambda))$ $\in\mathfrak{\tilde
{g}}_{+}$ at the element $\alpha(\lambda)=\lambda^{3}I+\lambda^{2}U\ +\lambda
V+Z$ \ then equals
\begin{equation}%
\begin{array}
[c]{c}%
\mathrm{P}_{+}\nabla\gamma_{2}^{(4)}(\lambda^{3}I+\lambda^{2}U\ +\lambda
V+Z)=\lambda^{4}I+2\lambda^{3}U+\\
\\
+\lambda^{2}(2V+U^{2})+\lambda(2Z+UV+VU)+\mathrm{M}(ZU+UZ)\ .
\end{array}
\label{M2.3}%
\end{equation}
From \ the commutator relationship (\ref{M1.12}) one obtains that the
following matrix equation
\begin{equation}%
\begin{array}
[c]{c}%
\frac{d}{dt}(\lambda^{2}U\ +\lambda V+Z)=[\lambda^{3}I+3\lambda^{2}%
U/2+\lambda(3Z/2-3U^{2}/8)-\\
\\
-U^{3}/16-3(ZU+UZ)/8\ ,\lambda^{2}U\ +\lambda V+Z]
\end{array}
\label{M2.3a}%
\end{equation}
holds for any parameter $\lambda\in\mathbb{C}.$ The trace-functionals
$H_{m}:=tr(\lambda^{3}I+\lambda^{2}U\ +\lambda V+Z)^{m/3},$ $m\in
\mathbb{Z}_{+},$ are all nontrivial and involutive with respect to the Poisson
bracket \ (\ref{M1.10b}) conservation laws of the Riemann type discrete matrix
dynamical system
\begin{align}
dU/dt  &  =-2[Z,V],\label{M2.3b}\\
& \nonumber\\
dV/dt  &  =-[Z,VU+UZ]+[UZ+ZU,V],\nonumber
\end{align}
easily following from \ (\ref{M2.3a}). Moreover, the discrete matrix dynamical
system \ (\ref{M2.3b}), as follows from is a completely integrable discrete
approximation of the corresponding limiting as $N\rightarrow\infty$ \ Riemann
type equations in partial derivatives
\begin{equation}
du/dt=\ -2v_{x},\text{ \ \ \ \ }dv/dt=uv_{x}-vu_{x}. \label{M2.3c}%
\end{equation}
Thus, \ concerning the limiting dynamical system \ (\ref{M2.3c}) one can
formulate the following proposition.

\begin{proposition}
\label{Prop_M2.2}The dynamical system \ (\ref{M2.3c}) allows for any
$N\in\mathbb{Z}_{+}$ the completely integrable \ matrix discretization
(\ref{M2.3b}), which allows the Lax type representation \ \ (\ref{M2.3a}).
\end{proposition}

As a simple consequence of Proposition \ref{Prop_M2.2}, the obtained above
limiting dynamical system \ (\ref{M2.3c}) is also Lax type integrable. More
applications of the devised above Calogero type discrete approximation
approach to different Lax type integrable nonlinear dynamical systems are
under preparation.

\section{Conclusion}

An observation, that the \ Calogero type discretization of the Heisenberg-Weyl
algebra is related to the Markov type splitting of the general Lie algebra
$gl(N;\mathbb{R}),N\in\mathbb{Z}_{+},$ proved to be both interesting and
useful for analytical constructing completely integrable discretizations of
the Lax type integrable nonlinear dynamical systems on functional manifolds.
They are represented as co-adjoint orbits on the Markov type co-algebras \ and
analyzed by means of the modern Lie-algebraic techniques. It is shown that a
set of conservation laws and the associated Poisson structure ensue as a
byproduct of the approach devised. Based on the \ Lie algebras
quasi-representation property the limiting procedure of finding the nonlinear
dynamical systems on the corresponding functional spaces is demonstrated.

\section{Acknowledgements}

Author is cordially thankful to Prof. Jan Cie\'{s}li\'{n}ski (Bia\l ystok
University, Poland) and Prof. Maxim Pavlov (Lomonosov's State University of
Moscow, Russian Federation) for friendly cooperation and fruitful discussions.
He also expresses cordial thanks to Prof. Francesco Calogero ("La Sapienza"
University of Roma, Italy) for \ interest in this work and useful comments.
The work was in part supported both by \ AGH\ University of Science and
Technology of Krakow, Poland, and the Scientific and Technological Research
Council of Turkey (TUBITAK/NASU- 110T558 Project) grants.

\bigskip

\bigskip
\end{document}